\documentclass{svjour3}
\smartqed
\usepackage[size=b5paper]{bxpapersize}
\usepackage[T1]{fontenc}
\usepackage{graphicx}
\usepackage{hyperref}
\usepackage{multirow}
\usepackage{url}
\usepackage{color}
\usepackage{xcolor}
\usepackage[nobreak]{cite}
\usepackage{xspace}
\usepackage{subfig}

\def\B{\allowbreak}

\newcommand{\Tool}[0]{\textsf{Ammonia}\xspace}

\journalname{Empirical Software Engineering}

\begin{document}

\title{Ammonia: An Approach for Deriving Project-specific Bug Patterns}

\author{Yoshiki Higo \and
            Shinpei Hayashi \and
            Hideaki Hata \and
            Meiyappan Nagappan
}

\institute{Yoshiki Higo \at
              Graduate School of Information Science and Technology, Osaka University\\
              1--5, Yamadaoka, Suita, Osaka 565--0871, Japan\\
              \email{higo@ist.osaka-u.ac.jp}
 \and
              Shinpei Hayashi \at
              School of Computing, Tokyo Institute of Technology\\
              Ookayama 2--12--1--W8--71, Ookayama, Meguro-ku, Tokyo 152--8550, Japan\\
              \email{hayashi@c.titech.ac.jp}
 \and
              Hideaki Hata \at
              Graduate School of Science and Technology, Nara Institute of Science and Technology\\
              8916--5 Takayama-cho, Ikoma, Nara 630--0192, Japan\\
              \email{hata@is.naist.jp}
 \and
              Meiyappan Nagappan \at
              Cheriton School of Computer Science, University of Waterloo\\
              200, University Avenue West Waterloo, Ontario Canada\\
              \email{mei.nagappan@uwaterloo.ca}
}

\date{}

\maketitle

\begin{abstract}
Finding and fixing buggy code is an important and cost-intensive maintenance task, and static analysis~(SA) is one of the methods developers use to perform it. 
SA tools warn developers about potential bugs by scanning their source code for commonly occurring bug patterns, thus giving those developers opportunities to fix the warnings~(potential bugs) before they release the software. 
Typically, SA tools scan for general bug patterns that are common to any software project~(such as null pointer dereference), and not for project specific patterns. 
However, past research has pointed to this lack of customizability as a severe limiting issue in SA. 
Accordingly, in this paper, we propose an approach called \Tool, which is based on statically analyzing changes across the development history of a project, as a means to identify project-specific bug patterns. 
Furthermore, the bug patterns identified by our tool do not relate to just one developer or one specific commit, they reflect the project as a whole and compliment the warnings from other SA tools that identify general bug patterns. 
Herein, we report on the application of our implemented tool and approach to four Java projects: \textsf{Ant}, \textsf{Camel}, \textsf{POI}, and \textsf{Wicket}. 
The results obtained show that our tool could detect 19 project specific bug patterns across those four projects. Next, through manual analysis, we determined that six of those change patterns were actual bugs and submitted pull requests based on those bug patterns. 
As a result, five of the pull requests were merged.

\end{abstract}

\section{Introduction}
\label{sec:introduction}

Software maintenance is a crucial activity during the development of any software product. There are several objectives to software maintenance, as evidenced by the thriving research community that has evolved around the International Conference on Software Maintenance and Evolution (ICSME). One of those objectives is to make sure that bugs in software are fixed. Past studies have shown that bugs can be costly and sometimes even cause harm to human life~\cite{zhivich2009security}. For those reasons, software practitioners use both preventive and corrective measures to address the issue of bugs. Some of the preventive techniques and analyses used by practitioners include testing~\cite{xiecite}, code review~\cite{rigby2008icse}, bug prediction~\cite{hall2012tse}, and static analysis (SA)~\cite{rahman2014icse,sadowski2015icse}, which are applied before the software is released to the end user. Corrective techniques and analyses include log file analysis~\cite{shang2015ese}, crash report analysis~\cite{kim2011tse}, and bug localization~\cite{wong2016tse}, among others, which are applied once the software is deployed to the end user. The bugs found will then be reported to the developers through bug reporting systems such as \textsf{Bugzilla} or \textsf{Jira}.

In this paper, we will focus on complementing one preventive technique - static analysis~(SA), which is a type of automated analysis that provides developers of the target software with warnings regarding potential bugs in their source code.  
The underlying idea behind SA tools is that there are some commonly occurring bugs across all software products~(even those written in different languages) and that such bugs often have identifiable patterns. 
For that reason, SA tools employ a set of rules (patterns) for commonly occurring bugs and scan the target source code to detect such patterns. 
For example, it is possible to automatically identify the code fragments where a bug like \textit{null pointer dereference}, which commonly appears in many software projects (including those written in different programming languages)~\cite{hoare2009}, can occur through a bug pattern. 
As a result, SA tools scan source code for such code fragments and report them as warnings to developers. 

Currently, there are a number of available SA tools. These include: \textsf{Splint}~\cite{splint2010}, \textsf{Cppcheck}~\cite{cppcheck2016}, \textsf{Clang Source Analyzer}~\cite{clang2015}, \textsf{FindBugs}~\cite{findbugs2015}, and \textsf{PMD}~\cite{pmd2015}.

Typically, the bug patterns in a software project are not just from a particular version of the target software project, but also cover the software development as a whole. However, while such bug patterns are beneficial, current SA tool databases do not contain any specific bug patterns that are part of a particular target software project, and researchers like Johnson et al.\ have previously pointed out that this lack of customizability is one of the reasons why SA tools are infrequently used~\cite{johnson2013icse}.

One of the reasons for the lack of project-specific bug patterns~(PSBPs) may be because there might not be any such patterns. 
However, Ray et al.\ found that developers make a non-trivial amount of similar changes in their software~\cite{ray2015msr}. 
Therefore, noting that there is empirical evidence that PSBPs do exist, we propose an approach called \Tool\ to identify PSBPs that are specific to particular software projects.

We identify the PSBPs by mining past bug-fix changes in the target software project. Our contributions in this paper are as follows:

\begin{itemize}
\item We propose an approach called \Tool, which complements ~(and does not replace), SA tools with bug patterns specific to a particular project.
\item We provide an implementation of our approach that is available for anyone to download and use.
\item We describe a case study where we apply our tool to four open source software systems and scan the latest versions of their source code to find PSBPs.
\item We evaluate the quality of the PSBPs identified in the case study systems and submit pull requests to fix the detected bugs.
\item We conclude with a candid discussion of where our methodology needs improvement so that future research can further develop our approach.
\end{itemize}

We begin by acknowledging that there are clone detection techniques and various SA tools that already exist. However, our approach combines these techniques and tools, along with change level analysis, in an effort to help developers and maintainers to find and fix commonly occurring bugs. To accomplish this, we overcame engineering challenges that helped scale the tool up for use in practical projects and not just toy examples. Hence, as an engineering research area, we believe that our contributions (bringing previous research ideas together, solving engineering challenges, building a working tool, and conducting a real-world empirical case study with fixed bugs), are highly relevant.

Note that the pattern identification portion of our proposed approach described in this paper is an enhanced version of our previous research~\cite{higo2012icsm}. 
However, the approach proposed herein includes the two major differences from the previous approach. Specifically:

\begin{itemize}
 \item The newly proposed approach includes code normalization and hash-based comparison to derive more appropriate change patterns. In contrast, source code lines are compared \textit{as they are} with the Unix \texttt{diff} command in the previous technique. The use of code normalization makes it possible to make a change pattern from code changes whose intrinsic contents are the same, even if their texts are different.
 \item Another enhancement is that the proposed approach considers bug-fix commits while the previous approach does not. Considering bug-fix commits makes it possible to focus on the most important changes and potentially reduces the number of false positives.
\end{itemize}

In this paper, we not only improve on our previous approach, we also build other tools such as a graphical user interface (GUI) tool that can be used by a developer to identify buggy code and find possible fixes for it. The resulting GUI is not simply a display of our results, it also provides users with the ability to filter the data as they seem fit. Currently, the GUI has filters that provide the following capabilities:

\begin{itemize}
 \item The ability to show only latent buggy code that matches with PSBPs, including given keywords in their commit logs.
 \item The ability to show only latent buggy code that matches with $n$-match PSBPs $n$ specified by a user.
 \item The ability to show only latent buggy code in files whose paths include specified keywords.
\end{itemize}

The first filter is useful when we want to concentrate on some specific types of buggy code. For example, ``race-condition'', ``null pointer'', or issue IDs would be useful keywords for this filter. 
The second filter is useful when we want to find latent buggy code efficiently because we empirically know that few-match PSBPs are more likely to be buggy code than many-match PSBPs. 
We assume that a user inputs 1 or 2 to use this filter. The third filter is useful when we want to concentrate on some specific files. For example, by using the filter, files under only a specific directory are shown to users.

The evaluation described in this paper was performed in a stricter manner. In this study, we made pull requests for each buggy code that we found using our proposed approach and submitted them to the software developers who then judged whether or not the pull requests were useful.

The rest of the paper is organized as follows: Section~\ref{sec:background} presents the background and definitions needed to understand our paper while Section~\ref{sec:approach} presents our approach and Section~\ref{sec:tool} provides a description of our tool. Section~\ref{sec:experiment} presents the case study that we carried out and its results, while Section~\ref{sec:discussion} presents a discussion of where our approach needs development~(so that future research can improve upon our work). Section~\ref{sec:relatedwork} presents our work within the context of other related work and Section~\ref{sec:threats} presents threats to validity in our study. Finally, Section~\ref{sec:conclusion} presents the conclusions of our study.

\section{Background and Definitions}
\label{sec:background}

In this section, we define the key terms behind our approach to identify PSBPs.

\subsection{Changes in Source Code}

When a bug is found as software is being used or tested, it is logged in a bug repository such as \textsf{Jira}/\textsf{Bugzilla}. Each such bug is then assigned to a developer who discusses it with colleagues and others, explores ways to fix it, and then submits a possible solution. This solution is then tested and reviewed by other developers. After successful testing and code review, the solution is committed to a source code repository such as \textsf{Git}/\textsf{Subversion}. Each such commit has two parts:

\begin{itemize}
 \item the before-change source code, which in the case of a bug is a chunk of problematic code, and
 \item the after-change source code, which in the case of a bug is a solution for the problematic code.
\end{itemize}

\begin{figure}[t]
 \centering%
 \includegraphics[width=.8\textwidth]{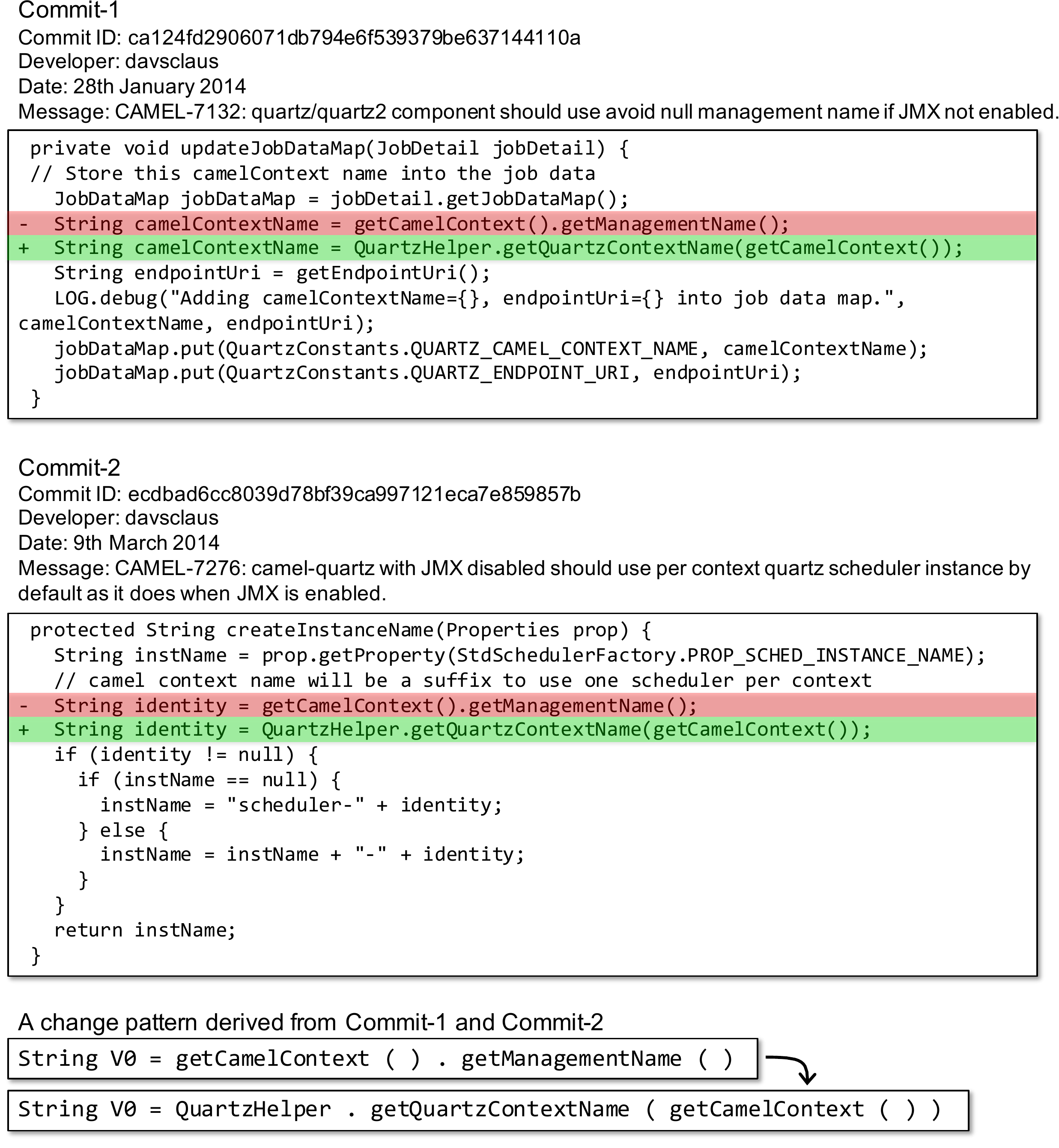}
 \caption{A change pattern in Apache Camel.}
 \label{fig:cpexample}
\end{figure}

The top of Figure~\ref{fig:cpexample} shows a concrete example of a commit that we extracted from \textsf{Apache Camel}. The line with prefix `--' is the before-change source code and the line with `+' is after-change source code.

\subsection{Change Patterns}
\label{sec:approach:cp}

The key idea behind our approach is that we mine all the commits in the entire development history of a specific project and identify change patterns among them in order to build a PSBP database. However, before we define what we mean by change patterns, let's first define the term \textit{code delta} as follows:

\begin{itemize}
 \item A \textbf{code delta} is a chunk of changed code. If a change is code addition, its chunk includes only after-change text. If a change is code deletion, its chunk includes only before-change text. If a change is code replacement, its chunk includes both before-change text and after-change text. In this research, we regard before-change text as an empty string in the case of code addition and after-change text is empty in the case of code deletion, respectively.
\end{itemize}

Then, we define a \textit{change pattern} as follows:

\begin{itemize}
 \item A \textbf{change pattern} is an abstract pattern that represents how source code was changed. A change pattern consists of code deltas whose both before-change text and after-change text are abstractly identical to one another. The reason why we abstract before-change and after-change texts is to disregard trivial differences among code deltas.
\end{itemize}

Figure~\ref{fig:cpexample} shows two commits from \textsf{Apache Camel}. In this figure, we can see that there are more than four commits that include the same code deltas. In total, the same code deltas occurred eight times in six different commits, and all of the code deltas form a single change pattern, as shown in the bottom of the figure.
If the commits from which the change pattern is extracted are bug fix commits, we can then call the change pattern a PSBP. In our approach, the history of a project is minded to extract a database of such PSBPs.

\section{Our Approach to Identify PSBPs}
\label{sec:approach}

In this section, we describe how we use our approach to determine PSBPs, which we call \Tool. There are three key phases in our approach:

\begin{figure*}[t]
 \centering%
\includegraphics[width=.98\textwidth]{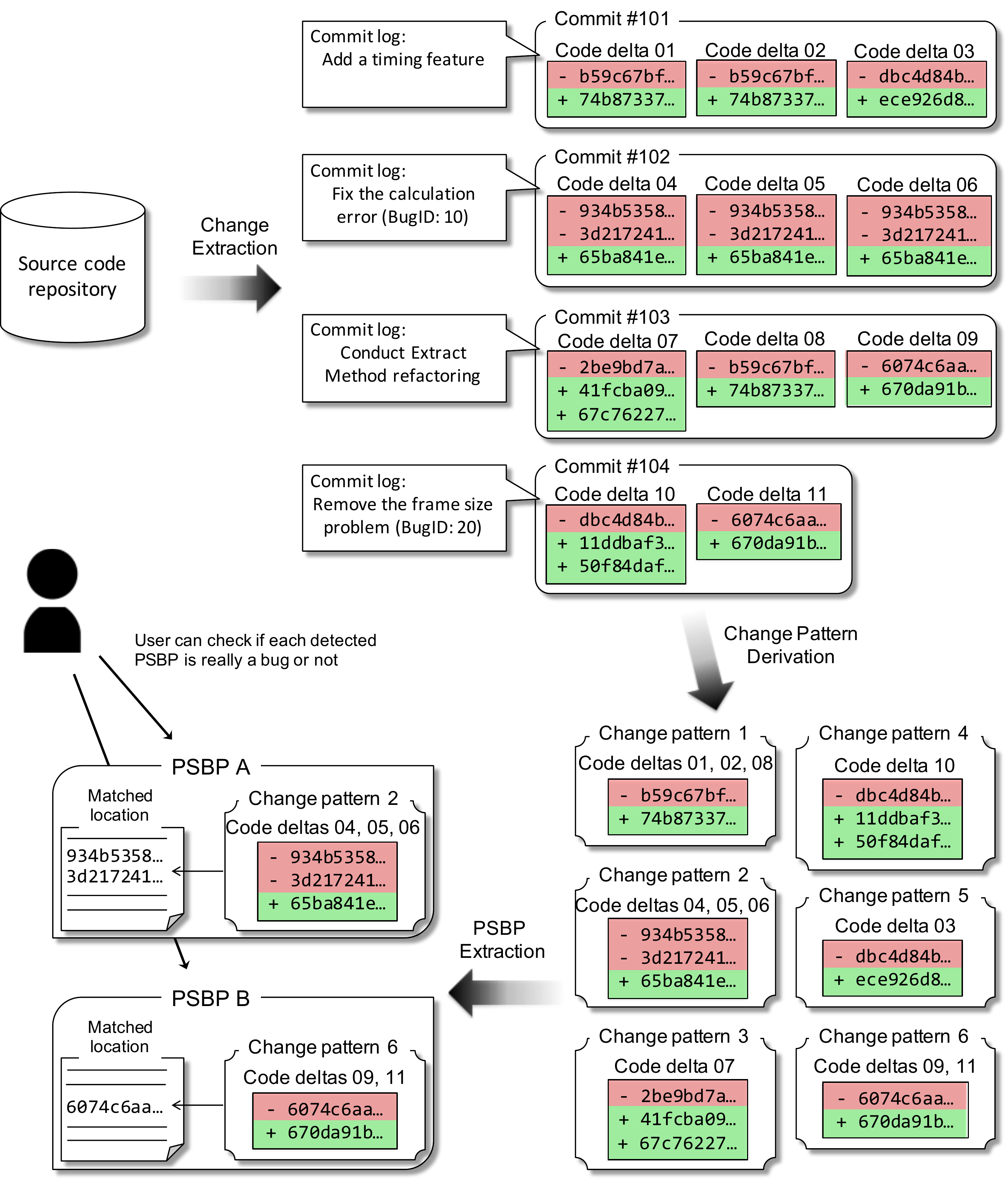}\caption{PSBP extraction process}
 \label{fig:ptoverview}
\end{figure*}

\begin{itemize}
 \item \textit{Change Extraction} -- For every commit in the development history of a particular project, we identify the actual changes made to the source code (i.e., the before-change and after-change texts) and then abstract them.
  \item \textit{Change Pattern Derivation} -- We then consider every abstracted change identified in the previous step, and group them to form change patterns.
 \item \textit{PSBP Extraction} - Then, based on certain conditions, extract PSBPs from the change patterns derived in the previous step. Developers can then determine if each of the extracted PSBPs is truly a bug-fix pattern.
\end{itemize}

Figure~\ref{fig:ptoverview} shows an overview of the proposed approach. In the following subsections, we describe each of the three phases.

\subsection{Change Extraction}

In the change extraction phase, we have three subprocesses:

\begin{enumerate}
 \item \textbf{Identify the source files changed in a given commit.} 
 A code repository contains not only source files, but also other kinds of files such as manual or copyright files. Such files are ignored, even if they are changed in the given commit, because our approach focuses solely on changes in the source files.
 \item \textbf{Abstracting the source files.} 
 We could limit our examination strictly to code deltas in the file, but if we only consider the code deltas, we face the following issues:

 \begin{itemize}
 \item Among other things, we do not know which token would be a variable and which would be a type.
 \item Only a part of a program statement is included in the code delta if the program statement is located within multiple lines of code of which only one line has been changed.
 \end{itemize}
 Hence, we abstract the entire source file from the revision before the commit and the same corresponding source file after the commit. To abstract the source files, we follow the five-step process shown in Figure~\ref{fig:ptcodeanalysis}. The example in this case is the source code before \texttt{Commit-1} in Figure~\ref{fig:cpexample}.

 \begin{figure}[t]
 \centering
 \includegraphics[width=.88\textwidth]{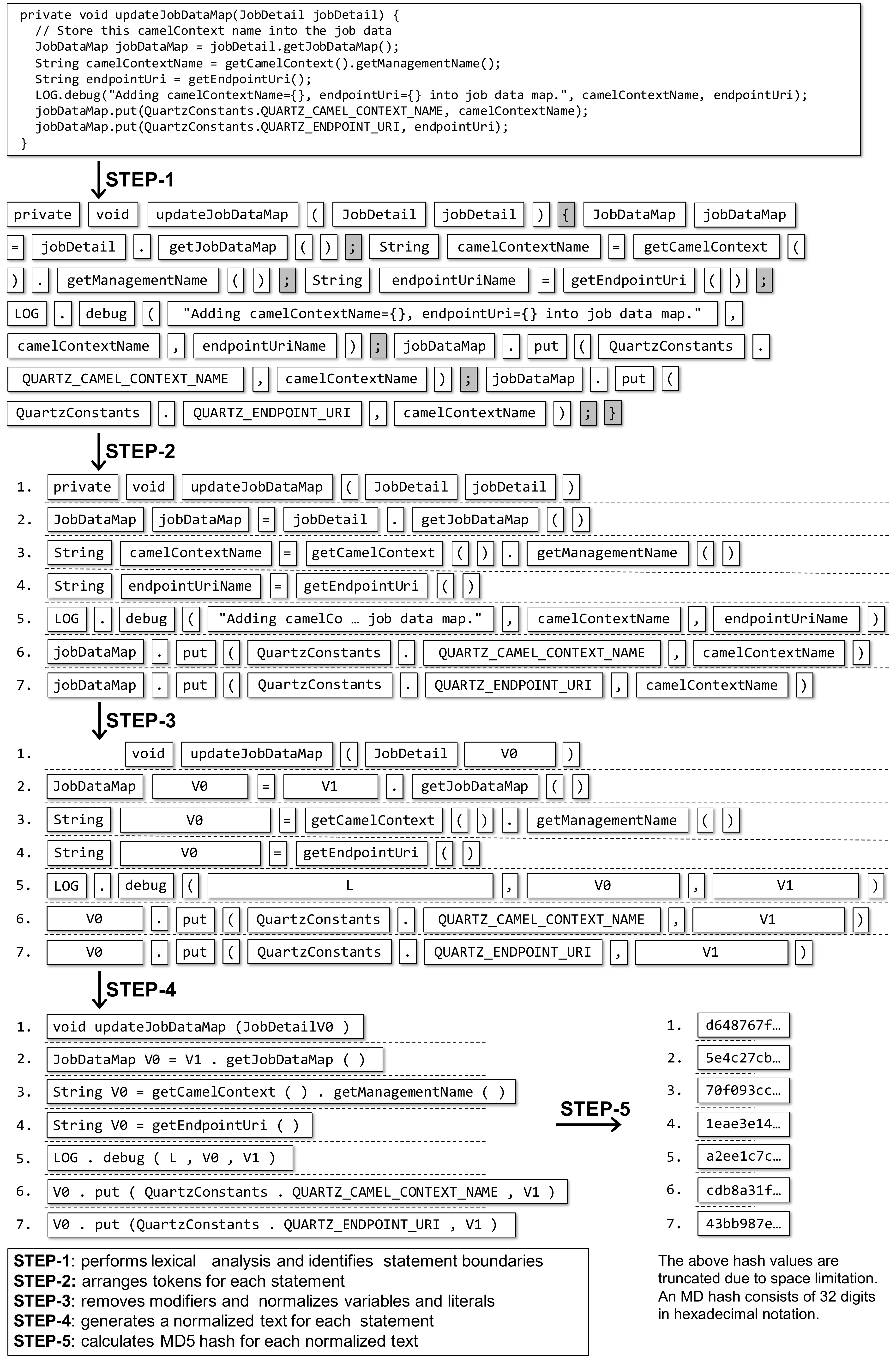}
 \caption{Technique for abstracting the source code files}
 \label{fig:ptcodeanalysis}
 \end{figure}
 
 \begin{figure*}[t]
 \centering
 \includegraphics[width=.9\textwidth]{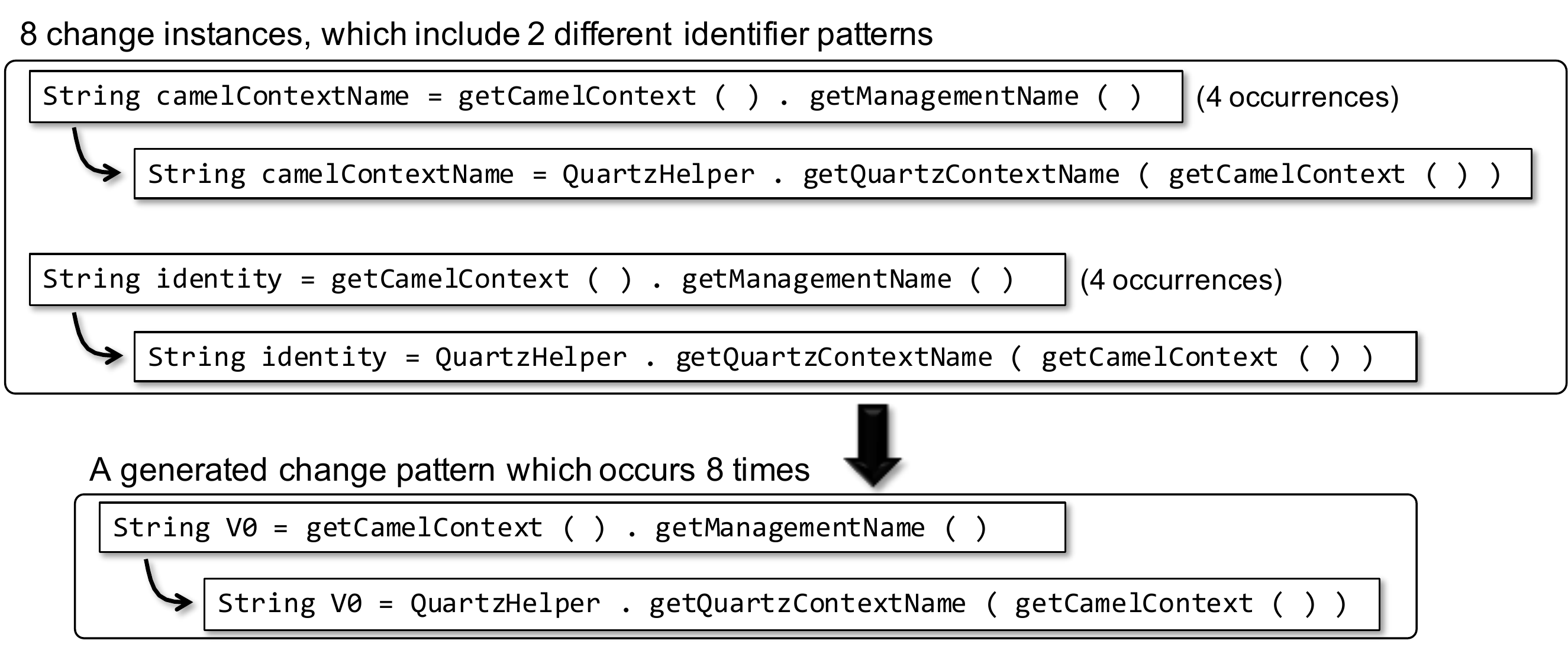}
 \caption{A change pattern derived from different texts}
 \label{fig:generatingcp}
 \end{figure*}

 \begin{description}
 \item[\textbf{STEP-1:}] We perform lexical analysis and identify statement boundaries. Three kinds of tokens, ``\texttt{;}'', ``\texttt{\{}'', and ``\texttt{\}}'' are used as statement boundaries.
 \item[\textbf{STEP-2:}] We then arrange tokens for each statement in a line.
 \item[\textbf{STEP-3:}] Next, we remove visibility modifiers such as ``\texttt{public}'' or ``\texttt{private}'' and normalize identifiers such as ``\textit{type names}'', ``\textit{primitive types}'', and ``\textit{variable names}''. 
 
Removing visibility modifiers is a design choice aimed at mitigating false positives, such as whether public/private should be added/removed for field declarations, which would cause our approach to point out a large number of false positives if not removed. It works by making it impossible to derive change patterns relating to adding/removing/changing visibility modifiers. However, at the same time, since removing visibility modifiers can reduce false positives, we decided it would be best to remove such visibility modifiers.

Variable names are normalized to ``\textit{V\#}''. The numbers of ``\textit{V\#}'' show the appearance pattern of variable names within a single statement. In each statement, the same numbers are assigned to the same names, and different numbers are assigned to different names. For example, three statements ``\texttt{a = a + 1;}'', ``\texttt{a = b + 1;}'', and ``\texttt{c = c + 1;}'' are normalized to ``\texttt{V0 = V0 + L;}'', ``\texttt{V0 = V1 + L}'', and ``\texttt{V0 = V0 + L}'', respectively. By normalizing code with this strategy, the same normalized text is generated from ``\texttt{a = a + 1;}'' and ``\texttt{c = c + 1;}'', but different normalized text is generated from ``\texttt{a = b + 1;}''.
We do not normalize method names because calls to different Application Program Interface (API) methods are very different semantically. 
We also normalize literals to \textit{L}. Another design choice we made was to normalize literals, because we empirically know that doing so can reduce false positives. An example of identifier normalization is shown at the bottom of Figure \ref{fig:cpexample}.
\item[\textbf{STEP-4:}] We generate a normalized line of text for each statement by concatenating tokens.
\item[\textbf{STEP-5:}] We calculate an MD5 hash for each normalized line of text.
\end{description} 
\item \textbf{Identify changes made by the commit.}
After abstracting the source files, we have a hash array for each source file. 
A hash array of each source file from before the commit is then compared to the hash array of the file from after the commit using the longest common subsequence (LCS) algorithm. 
By applying the LCS algorithm, we can identify deleted, added, and replaced hash values.
\begin{itemize}
 \item A hash subsequence deletion means a code deletion.
 \item A hash subsequence addition means a code addition.
 \item A hash subsequence replacement means a code replacement.
\end{itemize}
Note that the proposed technique utilizes only code deletion and code replacement because code addition cannot be utilized to identify code fragments that include latent bugs. 
\end{enumerate}

We repeat these three subprocesses for every commit in the entire development history of the project.

\subsection{Change Pattern Derivation}

In the change pattern derivation phase, we classify the extracted changes based on their before-change and after-change code deltas. If both the normalized before-change and after-change texts of any two given code deltas are the same, they are classified into the same group. Code fragment matching is performed with their MD5 hashes while both string and hash comparisons have similar performance. Figure~\ref{fig:generatingcp} shows the change pattern that we presented in Figure~\ref{fig:cpexample}. This pattern shows the importance of the identifier normalization in our proposed technique. The instances of this pattern include different variable names, \texttt{camelContextName} and \texttt{identity}. The same change occurred eight times in the development history of \textsf{Camel}, but includes two different identifier patterns. If the proposed technique did not include the code normalization, two different change patterns would have been derived. This is important because if a single change pattern is detected as two different patterns, it becomes more difficult to notice that the developers of \textsf{Camel} began using class \texttt{QuartzHelper} instead of method \texttt{getManagementName()}. Therefore, once we group every change identified in the previous phase, we have a collection of change groups, each of which is a change pattern, and thus a database of change patterns that are specific to a given project.

\subsection{PSBP Extraction}
\label{sec:approach:psbp}

Since the change patterns described in the last subsection are derived from all past changes, some of them are not related to fixing bugs. Therefore, in order to obtain change patterns that are more useful for finding latent problematic code in the latest version of the software project, we begin by filtering out change patterns that are not related to fixing bugs. More specifically, in our approach, we use the following two conditions:
Change patterns that satisfy both the conditions remain.
\begin{itemize}
 \item \textbf{Condition-1: change patterns related to bug-fix commits.}\\ Commits in the repository of the target software projects can be classified into bug-fix commits and other commits such as functional enhancement or refactoring. We only use change patterns in which at least one of their constituent changes have appeared in bug-fix commits. Our approach is designed to use the IDs of resolved and closed bug-related issues to identify bug-fix commits. If a given commit includes any of the bug-related issue IDs in its log messages, it is regarded as a bug-fix commit.
 \item \textbf{Condition-2: change patterns whose before-texts are different from the before-texts of any other change patterns.}\\ Although duplicated code fragments can be changed in different ways in version histories, in the case of bug-fix changes, we assume that the duplicated problematic code is changed in the same way. If two duplicated code fragments are changed in different ways, our proposed technique regards the two changes as two different change patterns. The two different change patterns share the same before-text, but their after-texts are different. We use only change patterns consisting of at least two changes and whose before-texts are different from the before-texts of all other change patterns.
 \end{itemize}

The remaining change patterns ((a) that are part of a bug-fix commit, and (b) have identical after-change texts for all the changes) are used to identify latent problematic (buggy) code. We identify such change patterns as PSBPs. Since the before-change text of the extracted patterns might be problematic code, we find code fragments in a given revision~(logically the latest revision, but potentially in any revision) that matches the before-change part of the change patterns. Matched code fragments with a PSBP are candidates of latent problematic code, and the after-change part of a PSBP is suggested to the developer as a possible fix for the buggy code.

We empirically know that there are some PSBPs whose before-change parts are matched with many code fragments in a given revision~(see Table~\ref{tbl:changepattern}). 
Single-match PSBPs are far fewer as seen in Table~\ref{tbl:changepattern} compared to all PSBPs. 
We did a manual analysis of many-match PSBPs (see Subsection~\ref{sec:discussion:many}). 
Since many-match code fragments are not latent problematic code, and since many-match PSBPs are rather useless, it is better to use the only PSBPs whose before-change text is matched with only a few code fragments in a given revision.

\section{Tool Description}
\label{sec:tool}

\begin{figure*}[t]
 \centering
 \includegraphics[width=.98\textwidth]{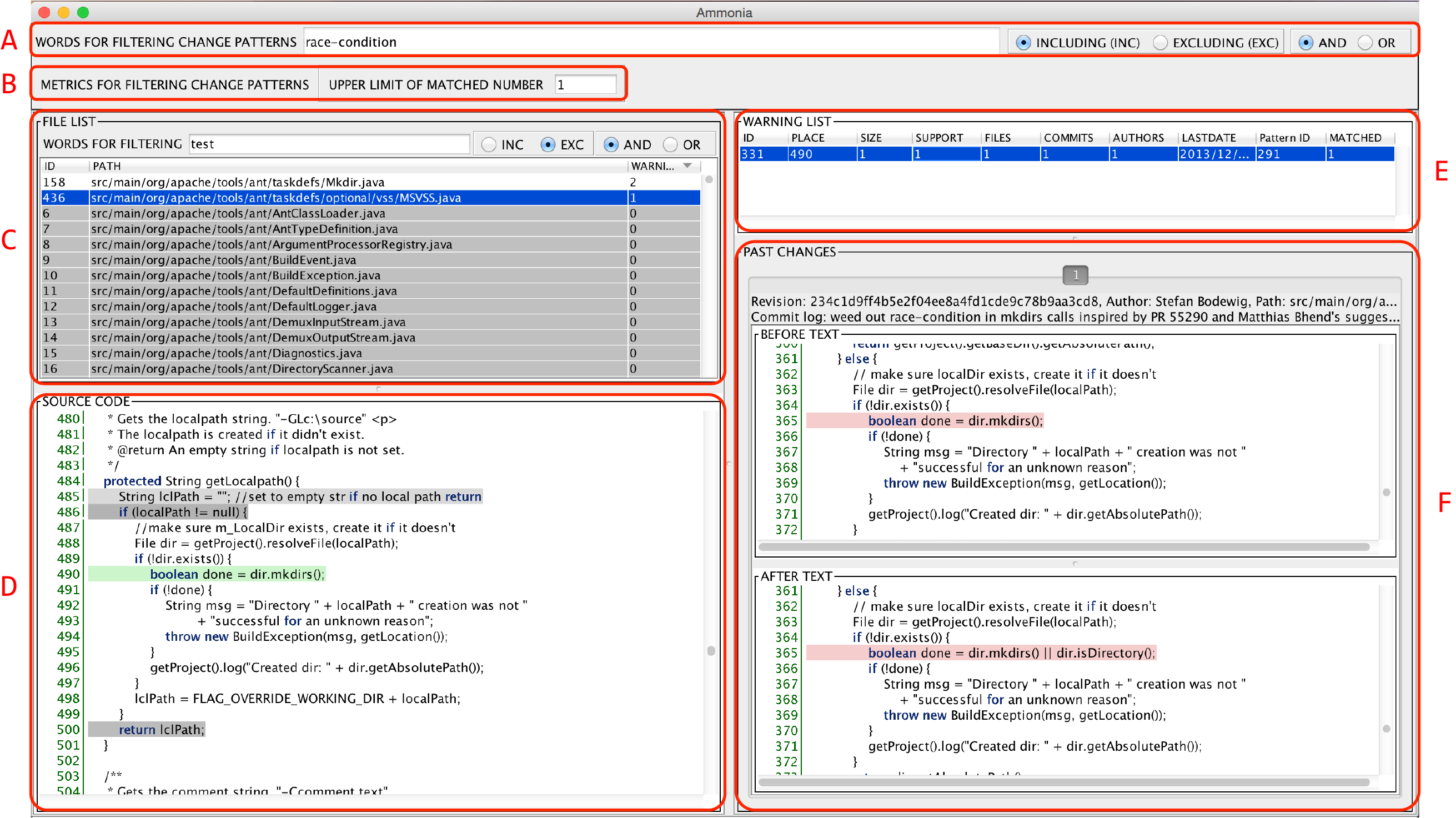}
 \caption{Tool snapshot}
 \label{fig:toolsnapshot}
 \end{figure*}

We have implemented a toolchain based on our proposed approach, which is shown in Figure~\ref{fig:ptoverview}. At this moment, our only target programming language is Java, but it will not be difficult to extend our proposed technique to other programming languages because it includes only lightweight source code analysis, such as a lexical analyzer. In cases where the toolchain supports another programming language, we simply need to implement a lexical analysis module and then specify tokens to be used as statement boundaries.

The first tool (a command-line tool) takes a software repository and finds change patterns, which are then stored in a structured query language (SQL) database. The second (GUI) tool, combines a version of a software project and the SQL database to first find latent buggy code from the version of source code. Next, it shows the matching results in a GUI window. Figure~\ref{fig:toolsnapshot} shows a snapshot of the second tool. A quick guide to using this tool is described below:

 \begin{itemize}
 \item Immediately after launching the GUI tool, source files in the target revision are listed in panel \textsf{C}, and all the other panels are empty. In \textsf{C}, each file has the number of matched code fragments in the given revision. The first action needed is selecting a file in \textsf{C}.
\item If a file in \textsf{C} is selected, panel \textsf{D} shows the source code of the file and panel \textsf{E} lists the set of PSBPs for the file. The second action is selecting a PSBP in \textsf{E}.
\item If a PSBP in \textsf{E} is selected, \textsf{D} automatically scrolls to the matched code of the selected PSBP and panel \textsf{F} shows past changes that were the reason for this suggestion. \textsf{F} provides before/after texts of code deltas included in the selected PSBP, the corresponding commit ID, and commit logs of the past changes. We assume that the users of this tool will investigate the PSBPs derived from our proposed approach with the information in panels \textsf{D} and \textsf{F}.
 \end{itemize}
 
The tool also has three filtering functions to remove inappropriately matched code suggestions.
 \begin{itemize}
\item Panel \textsf{A} is used to filter out change patterns. Code that matches with filtered-out change patterns is not suggested to the developers. In Figure~\ref{fig:toolsnapshot}, we are filtering out change patterns whose commits do not include the term ``race-condition''. Developers can use any keyword to search through the commit logs, and hence get any change patterns.
 \item Panel \textsf{B} is used to filter out change patterns based on the number of matches they have with the given revision. 
 It is expected that developers might want to examine change patterns that occur only once in the given revision~(an \textit{overlooked} bug), or change patterns that have several matches within the given revision~(a common bug).
 \item Panel \textsf{C} has a function to filter out files. 
 For example, when test files or tool-generated files are the targets of this filtering, we can remove them based on names included in their file paths. 
 In Figure~\ref{fig:toolsnapshot}, we are removing files that include ``test'' in their file paths.
 \end{itemize}

 To make it easier to identify useful/important change patterns from a huge number of such items, change patterns are characterized with some quantitative metrics in \textsf{E}.
 The following are the metrics used to characterize change patterns:
 \begin{itemize}
 \item \textsf{SIZE} is the number of statements in the before-text of the given change pattern.
 \item \textsf{FILES} is the number of distinct files where the given change pattern
 appears.
 \item \textsf{COMMITS} is the number of commits where at least an instance (an actual change) of the given change pattern appears.
 \item \textsf{AUTHORS} is the number of distinct authors that made commits where at least an instance of the given change patterns appears.
 \item \textsf{SUPPORT} is the number of instances included in a given change pattern. Note that \textsf{SUPPORT} and \textsf{COMMITS} are different because several instances of a change pattern can occur in the same commit.
 \item \textsf{MATCHED} is the number of code fragments in the target source code revision that match a particular change pattern, which is also used for the filtering function shown in Panel \textsf{B}.
 \end{itemize}
 
Our toolchain has been developed in Java, and is open to the public in GitHub.\footnote{\url{https://github.com/YoshikiHigo/NH3}} 
Since we wanted to determine if our tool could find a real-world bug before we carried out a full-fledged evaluation, we checked out the latest revision of \textsf{Apache Ant} issued on May 1, 2016, and then made a database of change patterns by using the command-line tool from the entire \textsf{Ant} history. 
Next, we launched the GUI tool with the latest revision and the database. 
The GUI tool showed many code fragments that matched with either change pattern because, at that time, we did not use Condition (b) (see Subsection~\ref{sec:approach:psbp}) and we did not restrict our search to single-match PSBPs, unlike the experiment described in Section~\ref{sec:experiment}. 
After investigating dozens of matched code fragments one-by-one, we found a code fragment that was very likely to be a bug in the file \textsf{src/\B main/\B org/\B apache/\B tools/\B ant/\B taskdefs/\B optional/\B vss/\B MSVSS.java}. 
We then contacted the developer via email, who had committed the code fragment, told us that the matched line of code was an overlooked part of his past bug-fix changes, and that he had fixed it immediately.%
\footnote{\url{https://github.com/apache/ant/commit/5c24a7}}
 The bugfix was then merged into the main branch of the \textsf{Ant} development.%
\footnote{\url{https://github.com/apache/ant/commit/fc0b2a}}

\section{Evaluating Our Approach}
\label{sec:experiment}

In this section, we evaluate the tool we implemented based on our approach~(\Tool) by applying it to four open source software projects. In the following subsections, we describe the open source software projects that we examined, the design of the evaluation, and the results obtained.

\begin{table}[b]
 \centering
 \caption{Case study subjects} \label{tbl:software}
 \begin{tabular}{lrrrrr}\hline
Project & \# bugs & First commit & Last commit & \# commits & \# bug-fix commits \\\hline
\textsf{Ant} & 2,007 & 4/Jan/2010 & 30/Jul/2016 & 673 & 208 \\
\textsf{Camel} & 2,618 & 19/Mar/2007 & 30/Jul/2016 & 23,861 & 4,687 \\
\textsf{POI} & 1,782 & 1/Feb/2002 & 29/Jul/2016 & 6,226 & 1,381 \\
\textsf{Wicket} & 2,654 & 23/Sep/2004 & 30/Jul/2016 & 23,363 & 2,621 \\\hline
 \end{tabular}
\end{table}

\subsection{Case Study Subjects}

Table~\ref{tbl:software} shows some information about the four software projects used in our evaluation. 
We provide information such as the first and last commit so that anyone wanting to replicate our evaluation results will be able to do so. 
All of the software projects are written in Java and are being developed in the Apache Software Foundation. 
We chose Java because our tool works on Java projects, but one could easily make changes to our tool~(which is available as an open source project) to work on software written in other languages as well. 
We also chose Apache Software Foundation projects since we wanted to use real-world projects and not toy examples. 
By examining real-world examples, we could also determine if our implementation has an adequate run time performance. 
The Git repositories for the four software projects are accessible via GitHub. 
We evaluated our tool on data from the four projects that has been uploaded before July 2016. 
As we can see from Table~\ref{tbl:software}, all projects have a similar number of bugs and each bug in the table has a corresponding report in the \textsf{Jira} reporting system. 

To apply our approach, we first need to determine whether or not each past commit is a bug-fix. Since the target software projects utilize \textsf{Jira}/\textsf{Bugzilla}, which are popular issue tracking systems, to manage issues on their systems, we collected the IDs of resolved and closed bug-related issues by using those systems. In this experiment, if a log message of a given commit includes any of the bug-related issue IDs, the commit is regarded as a bug-fix. 
The column of ``\# bugs'' of Table~\ref{tbl:software} includes the number of past bug fix commits that we collected.

\subsection{Evaluation Design}
\label{sec:experiment:design}

As described in Section~\ref{sec:tool}, our toolchain includes two tools. 
The first is a command line tool used to extract change patterns. 
We applied this tool to the code repositories of all four case study subjects in order to obtain a change pattern database for each of the four case study subjects. 
The second (GUI) tool takes the change pattern database obtained via the first tool and a target revision as input.
The target revisions that we chose for each case study subject are shown in Table~\ref{tbl:snapshots}. 
Note that there is no overlap between the chosen versions in Table~\ref{tbl:snapshots} (all in August 2016) and those in the input repositories (All up to July 2016). 
Using these two inputs, the GUI tool can identify latent buggy code in the chosen revision.
In this experiment, we use only single-match PSBPs to identify latent buggy code.

\begin{table}[b]
 \centering
 \caption{Target snapshots. The commit IDs are truncated. A whole commit ID consists of 40 digits in hexadecimal notation. For the four target projects, the seven digits are sufficient for identifying the target commits (git-log command works with the seven digits).} \label{tbl:snapshots}
 \begin{tabular}{lrrrr}\hline
Project & Commit ID & Commit date & \# files & LOC \\\hline
\textsf{Ant} & \texttt{1de4dfa...} & 7/Aug/2016 & 866 & 223,016 \\
\textsf{Camel} & \texttt{dc77701...} & 1/Aug/2016 & 4,949 & 277,111 \\
\textsf{POI} & \texttt{34a6732...} & 11/Aug/2016 & 2,216 & 431,853 \\
\textsf{Wicket} & \texttt{ba393ff...} & 20/Aug/2016 & 1,861 & 287,421 \\\hline
 \end{tabular}
\end{table}
   
\begin{table}[b]
 \centering
 \caption{Number of change patterns and single-match PSBPs found by our tool}
 \label{tbl:changepattern}
 \begin{tabular}{lrrrr}\hline
\multirow{2}{*}{Project} & \# all change & \# change patterns & \# change patterns & \# single-match \\
& patterns & satisfying (a) & satisfying (a) and (b) & PSBPs \\\hline
\textsf{Ant} & 3,975 & 644 & 30 & 1 \\
\textsf{Camel} & 73,802 & 9,851 & 1,573 & 7 \\
\textsf{POI} & 47,234 & 9,623 & 2,052 & 9 \\
\textsf{Wicket} & 55,272 & 4,317 & 532 & 2 \\\hline
 \end{tabular}
\end{table}

\begin{table}[b]
 \centering
 \caption{Manual investigation results for single-match PSBPs}
 \label{tbl:manualresults}
 \begin{tabular}{lrrrr}\hline
Project & \# PSBPs & Buggy & Non buggy & Unknown \\\hline
\textsf{Ant} & 1 & 1 & 0 & 0 \\
\textsf{Camel} & 7 & 3 & 2 & 2 \\
\textsf{POI} & 9 & 1 & 8 & 0 \\
\textsf{Wicket} & 2 & 1 & 1 & 0 \\\hline
Total & 19 & 6 (31.6\%) & 11 (57.9\%) & 2 (10.5\%) \\\hline
 \end{tabular}
\end{table}

\subsection{Results}
\label{sec:experiment:results}

In Column 2 of Table~\ref{tbl:changepattern}, we present the total number of change patterns extracted from each of the projects. For \textsf{Ant}, we only use commits data after January 1st, 2010 to derive change patterns because prior to January 2010, \textsf{Ant} underwent significant design alterations that resulted in method name, logging, and exception handling changes. As a result, only 3,975 change patterns were derived from \textsf{Ant}, with the other three case subjects having at least one order of magnitude more change patterns.

As explained in Subsection \ref{sec:approach:psbp}, we use change patterns satisfying two conditions: (a) change patterns whose changes occurred in bug fix commits at least once, and (b) change patterns whose after-change texts are the same for all the changes. Columns 3 and 4 of Table~\ref{tbl:changepattern} shows the number of change patterns satisfying (a) and the number of change patterns satisfying both (a) and (b). The change patterns satisfying both (a) and (b) are used to identify PSBPs.

In Column 5 of Table~\ref{tbl:changepattern}, we present the number of PSBPs found from each of the projects. Those numbers are PSBPs that have only a single match in the chosen revision of the case study subjects. In this experiment, we used only single-match PSBPs because, as described in Subsection~\ref{sec:approach:psbp}, we know that as the number of code fragments a PSBP matches with increases, the less harmful those matched code fragments are.

Table~\ref{tbl:manualresults} shows the results of the manual analysis we had carried out for each single-match PSBP before submitting pull requests. The following is an explanation for the last three columns of the table.

\begin{itemize}
 \item \textbf{Buggy.} The number of matched code fragments that we determined as having the same bugs as the before-change text in the PSBP.
 \item \textbf{Non buggy.} The number of matched code fragments that~(based on manual analysis) we did not regard as having the same bug as the before-change text in the PSBP.
 \item \textbf{Unknown.} The number of code fragments that we were not able to make any conclusions about, even after careful manual investigation. 
 The reason for this is because we are neither the developers nor experts in the case study systems that were examined. 
 While it is likely that the relevant system developers could comment better on these uncertain code fragments, we did not want to waste their time by asking them for commits. 
 Therefore, even though this remains an issue, we do not consider those code fragments to be useful and removed them from consideration in order to prevent distorting our results.
\end{itemize}

In the judgment process, we first attempted to determine if each matched code fragment should be considered a false positive. If we were able to find a reason, we confirmed it as a false positive and regarded the code fragment as \textit{Non buggy}. If we were not able to find any reason to regard it as a false positive, and we considered it likely that the code fragment included the same bug as the PSBP, we regarded it as \textit{Buggy}. In cases where we were unable to find reasons but did not consider it likely that the code fragment included the same bug, we regarded it as \textit{Unknown}. The reasons used in this identification process are discussed in Section~\ref{sec:discussion}.

In total, 19 code fragments were suggested as potential latent bugs. Our manual analysis then determined that six~(approximately 6/19=31.6\%) of the matched code fragments were actual bugs. While a precision level of around 30\% seems low, note that the number of matched code fragments that remained listed as latent bugs after the filtering provided by our tool dropped to just 19. In other words, from thousands of change patterns, our approach identified only single-digit PSBPs per project~(unlike the warnings from other SA tools that number in the hundreds or thousands). Hence, even though the precision level is low, since the total number of PSBPs is small, developers should be able to check each of them manually.

Liu et al.\ experimented with 730 OSS projects with FindBugs~\cite{liu2018tse} and found 16,918,530 distinct code violations, but the developers removed only 88,927 out of them. In other words, the number of removed violations was only 0.5\%, which is much less than 31.6\% removed via the use of our process.

\newcommand{\PR}[6]{#1 & #2 & #3 & #4 & #5 & #6}
\newcommand{\PRS}[6]{#2 & #3 & #4 & #5 & #6}
\newcommand{\CNG}[2]{{\fontsize{7pt}{0cm}\selectfont\MP{6.7cm}{%
 \vspace{0.3em}%
 \texttt{#1}\vspace{-0.6em}%
 \begin{flushright}$\hookrightarrow$ \texttt{#2}\end{flushright}%
 \vspace{-0.2em}%
 }}}
\newcommand{\CNGS}[2]{{\fontsize{7pt}{0cm}\selectfont\MP{6.7cm}{%
 \vspace{0.2em}%
 \texttt{#1} $\to$ \texttt{#2}%
 \vspace{0.2em}%
 }}}
\newcommand{\MP}[2]{\begin{minipage}{#1}#2\end{minipage}}
\newcommand{\DMY}[3]{\MP{2em}{#1/#2\\/#3}}
\newcommand{\DATE}[2]{\MP{6em}{\vspace{0.2em}#1\\~~~~~~~to \\#2\vspace{0.2em}}}
\newcommand{\reducedstrut}{\vrule width 0pt height .9\ht\strutbox depth .9\dp\strutbox\relax}
\newcommand{\Mod}[1]{%
 \begingroup
 \setlength{\fboxsep}{0pt}%
 \colorbox{black!25}{\reducedstrut#1\/}%
 \endgroup
}

\begin{table}[b]%
 \centering%
 \caption{Pull requests for bug-related issues}\label{tbl:pullrequests}
 {\scriptsize\renewcommand\arraystretch{1.0}\tabcolsep=0.4em\begin{tabular}{lrlcrl}\hline
 Project\hspace{-0.4em} & ID & Status & Dates & \textsf{SUPPORT} & Suggested change \\\hline
 \PR{\textsf{Ant}}{20}{Rejected}{\DMY{3}{Feb}{2011}}{57}%
 {\CNG{for (int i = 0; i < \Mod{children.size()}; i++)}%
 {\Mod{final int size = children\B.size() ; }\\for (int i = 0; i < \Mod{size}; i++)}}
 \\ \hline
 \PR{\textsf{Camel}\hspace{-0.3em}}{1108}{Merged}{\DMY{26}{Apr}{2013}}{3}%
 {\CNG{byte[] bytes = context\B.getTypeConverter()\\~~.\Mod{convertTo}\B(byte[]\B.class, in)}%
 {byte[] bytes = context\B.getTypeConverter()\B.\Mod{mandatoryConvertTo}\B(byte[]\B.class, in)}}
 \\ \cline{2-6}
 \PR{}{1137}{Merged}{\DMY{28}{Jan}{2014}}{2}%
 {\CNG{String camelContextName = \Mod{getCamelContext()\B.getManagementName}()}%
 {String camelContextName = ~~~~~~~~~~\\\Mod{QuartzHelper\B.getQuartzContextName}\B(\Mod{getCamelContext()})}}
 \\ \cline{2-6}
 \PR{}{1142}{Merged}{\DMY{28}{Aug}{2015}}{3}%
 {\CNG{Exchange exchange =\\\Mod{new DefaultExchange}\B(\Mod{this, }pattern)}
 {Exchange exchange =~~~~~~~~~~~~~\\\Mod{super\B.createExchange}\B(pattern)}}
 \\\hline
 \PR{\textsf{POI}}{36}{Merged}{\DMY{1}{Nov}{2015}}{3}%
 {\CNG{XSSFPivotTable pivotTable = sheet\B.createPivotTable\B(\\~~new AreaReference\B("A1:D4"),\\~~new CellReference\B("H5"))}%
 {XSSFPivotTable pivotTable = ~~~~~~~~~~~~~~~~\\sheet\B.createPivotTable\B(%
 new AreaReference\B(\\"A1:D4"\Mod{, SpreadsheetVersion\B.EXCEL2007}),\\new CellReference\B("H5"))}}
 \\\hline
 \PR{\textsf{Wicket}\hspace{-0.3em}}{179}{Merged}{\DMY{15}{Jun}{2010}}{2}%
 {\CNG{response\B.\Mod{setContentLength}\B(\Mod{(int) }length)}
 {response\B.\Mod{addHeader}~~~~~~~~~~~~~~~~~~~~~~~~\\(\Mod{"Content-Length", Long\B.toString\B(}length\Mod{)})}}
 \\\hline
 \end{tabular}}
\end{table}

After the manual investigation that had been conducted in order to confirm if the latent bugs that our tool identified were actually bugs, we submitted pull requests for six of them. 
Table~\ref{tbl:pullrequests} presents the details about all six of the pull requests. The ID column presents the pull request ID for each project and can be used to see the pull request on GitHub.\footnote{\url{https://github.com/apache/{ant,camel,poi,wicket}/pull/<ID>}}
We also present the status of the pull requests and when their status was last changed, the \textsf{SUPPORT} value for the change pattern associated with each pull request (as this signifies the number of changes in the past that has had the same bug fixed), and the actual change associated with each pull request.

From the results, it can be seen that five (83.3\%) of the six pull requests have been merged and one pull request in \textsf{Ant} was rejected. The developer rejected the last pull request because it would introduce a new bug to \textsf{Ant}.\footnote{\url{https://github.com/apache/ant/pull/20}} The suggested change was a micro-optimization aimed at improving \textsf{Ant}'s performance by avoiding multiple invocations of \texttt{size()}, which has occurred 57 times in the past. However, in this case, \texttt{children} can be added dynamically. Consequently, optimizing the loop by replacing \texttt{children.size()} with a variable would break \textsf{Ant}'s behavior.

We also ran \textsf{PMD}, which is a popular SA tool, on the same snapshot of the four systems to which we applied \textsf{Ammonia} and found that \textsf{PMD} was not able to find latent buggy code for any of the 19 single-match PSBPs including the ones which we submitted as pull requests and were accepted by the developers. 
Thus we can see that Ammonia can find issues that are not detected by a static analysis tool like PMD. 

\noindent \textbf{Evaluation Summary:} \textit{Our tool was able to successfully extract PSBPs from the case study subjects and about 31.6\%~(six out of 19) of the PSBPs resulted in the identification of actual bugs in cases where only single-match PSBPs were used. 
Note that like any current bug detection technique, we were unable to find all possible bugs, so it is impossible to measure recall. All we can measure is precision and our current effectiveness. Nevertheless, we successfully confirmed that about 31.6\% of the identified PSBPs could be used to fix bugs in four case study subjects.}

\section{Discussion}
\label{sec:discussion}

Herein, we discuss the results that we obtained in the experiment.
First, Subsection~\ref{sec:discussion:reasons} describes the reasons why we judged the matched code fragments as \textit{Non buggy}.
Second, we show the results of another experiment in the case that we used not only the bug-related issue IDs but also all the issue IDs.
Third, we show some examples of the matched code fragments that were found with many-match PSBPs while we only investigated single-match PSBPs in the experiment of Section~\ref{sec:experiment}.

\subsection{Reasons Why We Judged the Matched Code Fragments as Non Buggy}
\label{sec:discussion:reasons}

From Table~\ref{tbl:manualresults}, we can see that about 31.6\% of the code fragments that matched with the PSBPs are bugs. While this is level of precision is quite good~(in comparison to SA tools~\cite{ayewah2010issta,liu2018tse}), it still means that about 57.9\% of the matched patterns were false-positives. Herein, we explain the reasons why we judged the matched code fragments as \textit{Non buggy}, focusing on three particular reasons we identified in the judgment process of the experiment.

\begin{itemize}
\item \textbf{Accidental coincidence.} There were cases where the text in the change corresponds to a method call, and the name of the method is very generic, like \textit{size()}. Hence, we initially matched a code fragment with a method that has the same name as the PSBP, but on further perusal found that the invoked methods are indeed very different. Since we do not abstract method names in our approach~(see Section~\ref{sec:approach}), we will avoid any more such instances.
\item \textbf{Mismatched context.} The context of a matched code fragment was different from the context of code fragments where changes included in a given PSBP occurred. For example, there are class $A$ and its subclasses $B$ and $C$. The PSBP was derived from changes that occurred in $B$ and $C$, but the matched code fragment is in $A$. Accordingly, we concluded that applying the same change to the parent class was inappropriate.
\item \textbf{Extract method.} The matched code fragment was refactored via extract method refactoring, but the before-change text of the given change pattern in this case was a multi-line code chunk, and its after-change text was a method invocation. Hence, the matched code fragment was actually the body of the extracted method.
\end{itemize}

\begin{table}[b]
 \centering
 \caption{Classification of \textit{Non buggy} code fragments} \label{tbl:rq2reason}
 \begin{tabular}{lrrr}\hline
\multirow{2}{*}{Project} & Accidental & Mismatched & Extract \\
 & coincidence & context & method \\\hline
\textsf{Ant} & 0 & 0 & 0  \\
\textsf{Camel} & 1 & 2 & 0 \\
\textsf{POI} & 0 & 6 & 2 \\
\textsf{Wicket} & 0 & 1 & 0 \\\hline
 \end{tabular}
\end{table}

Table~\ref{tbl:rq2reason} shows the number of \textit{Non buggy} code fragments that were classified based on each of the three reasons above. For all case study subjects except \textsf{Ant}, mismatched context was the biggest reason for false positives. Since our proposed approach does not consider the context surrounding the matched code, many \textit{Non buggy} code fragments were misidentified due to this reason.

For \textsf{POI}, refactored code are matched as well. Although it is possible to exclude them automatically if we can identify and track refactoring changes~\cite{mahouachi2013gecco,prete2010icsm,xing2006wcre}, the time required to mine software repositories will be much longer if we use such techniques. In other words, it is a trade-off between accuracy and performance.

For \textsf{Camel}, there was one case of accidental coincidences. Since our proposed approach employs text-based rather than entity-based matching with semantic analysis, we expected such false positives to occur, but we believe that the number of code fragments identified due to this reason is small enough that developers can easily determine that those code fragments are \textit{Non buggy}.

\begin{figure}[t]
 \centering
 \subfloat[Introducing a local variable to stop double invocations (16 matches)]{
 \includegraphics[width=0.72\textwidth]{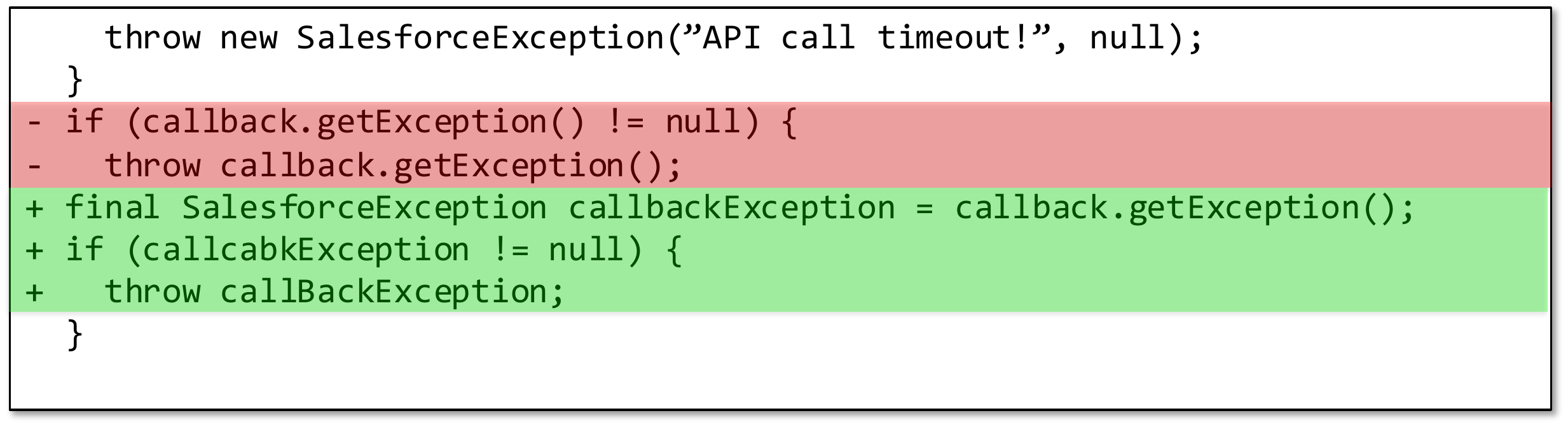}
 \label{fig:refactoring1}
 }\\
 \subfloat[Introducing a method invocation for finalizations (13 matches)]{
 \includegraphics[width=0.72\textwidth]{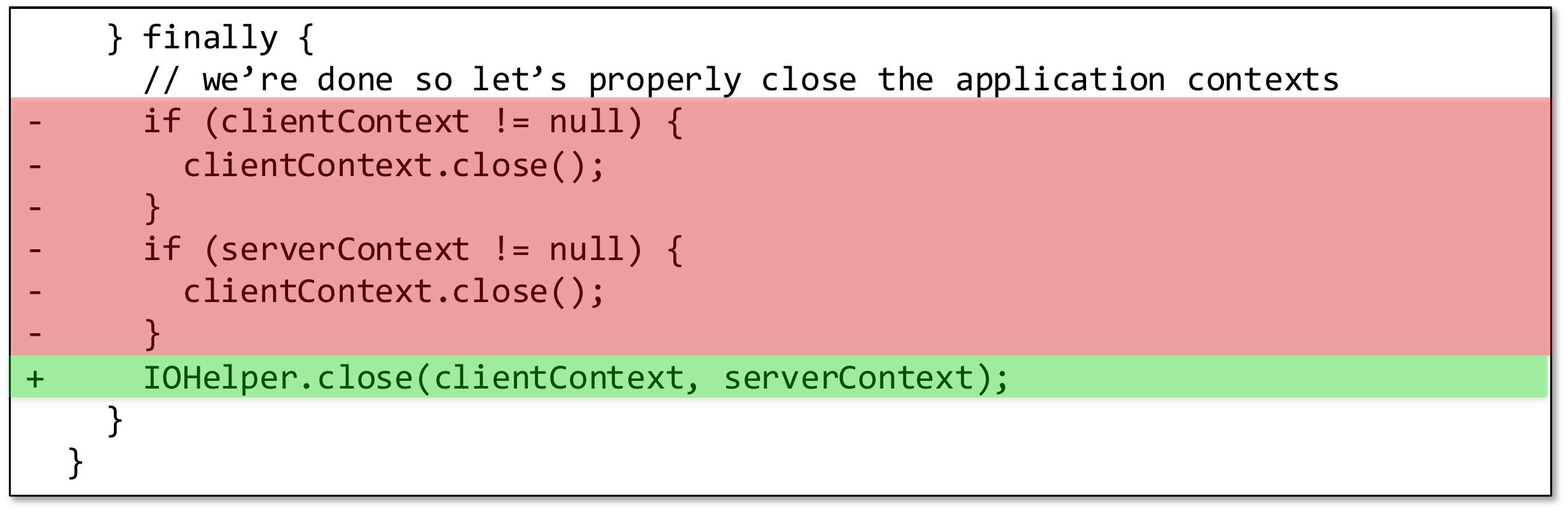}
 \label{fig:refactoring2}
 }\\
 \subfloat[Using \texttt{toArray} instead of \texttt{copyInto} (8 matches)]{
 \includegraphics[width=0.72\textwidth]{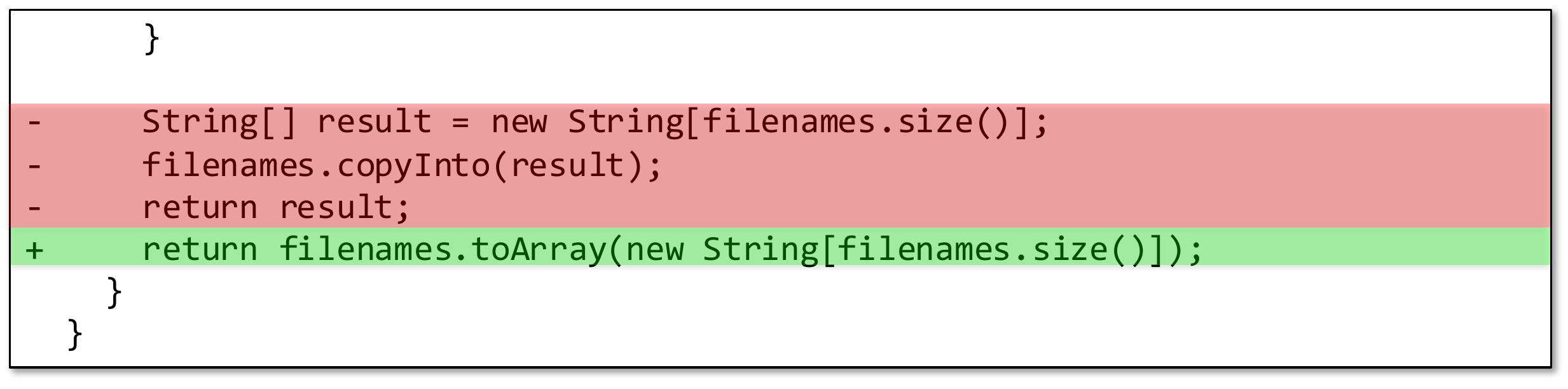}
 \label{fig:refactoring3}
 }
 \caption{Micro refactoring examples}
 \label{fig:refactorings}
 \end{figure}
 
\subsection{Using Non Bug-Related Issue IDs}
\label{sec:discussion:nonbug}

\begin{table}[b]%
     \centering%
     \caption{Pull requests for non bug-related issues}\label{tbl:pullrequests2}
     {\scriptsize\renewcommand\arraystretch{1.0}\tabcolsep=0.6em\begin{tabular}{rllrl}\hline
     ID & Status & Dates & \hspace{-3em}\textsf{SUPPORT} & Suggested change \\\hline
     \PRS{\textsf{Camel}\hspace{-0.3em}}{1134}{\MP{2em}{Reverted}}{\DATE{13/Mar/2011}{28/Mar/2011}}{47}%
     {\CNG{\Mod{if (logger\B.isTraceEnabled()) \{ }\\~~logger\B.trace("runningAllowed() -> "\Mod{ +} answer)\Mod{; \}}}%
     {logger\B.trace\B("runningAllowed() -> \Mod{\{\}}"\Mod{,} answer)}}
     \\ \cline{1-5}
     \PRS{}{\multirow{2}{*}{1135}}{\multirow{2}{*}{Merged}}{12/Feb/2015}{2}%
     {\CNG{return toDOMSource\B(source, \Mod{(Exchange) }null)}%
     {return toDOMSource\B(source, null)}}
     \\ \cline{3-5}
     \PRS{}{}{}{12/Feb/2015}{3}%
     {\CNG{return toDOMDocument\B(source, \Mod{(Exchange) }null)}%
     {return toDOMDocument\B(source, null)}}
     \\ \cline{3-5}
     \PRS{}{1136}{Merged}{22/Nov/2014}{7}%
     {\CNG{messageEvent\B.getChannel()\B.write(response)}%
     {messageEvent\B.getChannel()~~~~~~~~~~~~~~~~\\.write(response)\B\Mod{.syncUninterruptibly() ;}\\
     \Mod{messageEvent\B.getChannel()\B.close()}\phantom{ ;}}}
     \\ \cline{3-5}
     \PRS{}{1140}{Merged}{\DATE{28/Oct/2010}{10/Nov/2010}}{2}%
     {\CNG{hostName = \Mod{InetAddress\B.getLocalHost()\B.getHostName}()}
     {hostName = \Mod{InetAddressUtil\B.getLocalHostName}()}}
     \\ \cline{3-5}
     \PRS{}{1141}{Merged}{28/Jan/2010}{2}%
     {\CNG{return "sendTo(" + destination\\~~\Mod{ + (pattern != null ? " " + pattern : "")} + ")"}%
     {return "sendTo(" + destination + ")"}}
     \\\hline
     \end{tabular}}
    \end{table}

We only used bug-related issue IDs to identify bug-fix commits in the experiment; however, we consider using non bug-related issue IDs is also useful.
As an extra experiment, we extracted PSBPs from \textsf{Camel} by regarding commits whose message include ``\texttt{CAMEL-[0-9]+}'' as bug-fix commits.
As a result, we detected 56,563 change patterns satisfying (a), 4,163 change patterns satisfying both (a) and (b), and 133 single-match PSBPs, respectively.
In the experiment, we found seven single-match PSBPs from \textsf{Camel} with bug-related issue IDs, which means 126 single-match PSBPs were derived from non bug-related issue IDs.
We made pull requests from five out of the 126 single-match PSBPs and four of them were merged by the developers.
Table \ref{tbl:pullrequests2} shows the pull requests.
The code changes are for deleting an unnecessary casting, adding a \texttt{close} method invocation after data sending processing, using a better API, and simplifying a text generation.
The proposed technique was able to suggest such non bug-fix changes in addition to bug-fix changes.
Thus, we can use all issue IDs instead of bug-related issue IDs but then the false positives are going to increase because PSBPs derived from all issue IDs are suggesting changes other than bugfixing.
We cannot submit pull requests for all 126 single-match PSBPs because GitHub bans people who try to submit such large number of automated pull requests~\cite{carlson2019oss}.

\subsection{Finding Code Fragments without the Single-Match Limitation}
\label{sec:discussion:many}

We limited the number of matched code fragments to 1 in the experiment. To see the impact of this limitation, we also searched for code fragments without the limitation. 
As a result, 45, 631, 940, and 66 code fragments were matched to PSBPs for the four target software products without the limitation. 
We then manually investigated dozens of the code fragments and we found that matched code fragments are micro refactoring opportunities rather than latent buggy code. We show some examples of this in Figure~\ref{fig:refactorings}. For example, in Figure~\ref{fig:refactorings}\subref{fig:refactoring1}, we can see a change pattern that introduces a temporary variable to avoid invoking \texttt{getException} twice. This change pattern matches 16 code fragments. In Figure~\ref{fig:refactorings}\subref{fig:refactoring2} we see a change pattern that is used to simplify the finalization code. Here, 13 code fragments were matched to this change pattern. Meanwhile, Figure~\ref{fig:refactorings}\subref{fig:refactoring3} shows a change pattern that replaces the \texttt{copyInto} invocation with \texttt{toArray} invocation in order to make the code simpler. While we found many refactoring opportunities with many-match PSBPs, we think that it is difficult to evaluate the refactoring opportunities that were found. Bug-fix changes are clearly evaluated by checking whether or not the code change can fix the bug, even though there is neither a generic nor strict standard that can be used to evaluate micro refactorings. It is generally said that the size and complexity of the code are used as a standard, but in case of micro refactorings, there are not many differences in such values between before and after code changes. 
Multi-match PSBPs may be studied further as a way to identify micro-refactorings. But that is out of the scope of this work.

\section{Related Work}
\label{sec:relatedwork}

Several related studies influenced our approach. In this section, we divide them into the following subsections:

\subsection{Empirical Studies on SA Tools}

Ayewah and Pugh reported the results of an extensive review of \textsf{FindBugs} warnings in Google's code base~\cite{ayewah2010issta}. Although many current SA tools can find problems cheaply, some detected bug patterns do not accurately capture their developers' concerns. They also found that developers overvalue some severe bug patterns that are rarely feasible in practice, and yet undervalue subtle bug patterns that are often harmless, but which can cause serious problems that are hard to detect. Their study motivated us to not just examine general bug patterns captured in SA warnings, but also to look for PSBPs.

Rahman et al.\ compared SA tools~(\textsf{FindBugs}, \textsf{Jlint}, and \textsf{PMD}) on the context of defect prediction by using historical data \cite{rahman2014icse}. 
The reason for the comparison is that all three products are aimed at finding and removing defects efficiently and accurately. 
They reported that they have comparable benefits, and that SA tools can be enhanced using the information obtained from defect predictions. These findings motivated us to use historical data in our approach to finding bug patterns.

Avgustinov et al.\ tracked SA warnings over the revisions of various programs and investigated their developers' characteristics of introducing and fixing typical warnings in those program \cite{avgustinov2015icse}. From their experimental study of several open source projects written in Java, C++, Scala, and JavaScript, they captured the coding habits of individual developers. Their work was similar to this study in that we also analyze histories to capture some patterns, but we do not limit patterns to just SA warnings. Instead, we investigate all bug-related code changes within a given project.

\textsf{Tricoder} is a program analysis platform at Google \cite{sadowski2015icse} that can be used by developers to evaluate warnings, which can then result in accuracy improvements. Similar to their work, we also customize the bugs that we identify to a specific project. However, unlike them, we use development histories and do not start from the warnings in a SA tool. Additionally, we also provide possible fixes for the bugs detected.

\subsection{Empirical Studies on Source Code Changes}

Some empirical studies of source code evolution examined the nature of changes. For example, Nguyen et al.\ studied the repetitiveness of code changes~\cite{ray2013ase}. They considered changes as repeated if they matched other changes that have occurred in the past and found a high level of repetitiveness for small size changes. Regarding bugfix changes, they concluded that cross-project repetitiveness is higher than within projects, and that the repetitiveness of small size changes in bug fixing is higher than that of general changes. Meanwhile, Barr et al.\ studied the plastic surgery hypothesis, which posits that changes to a code repository have snippets that already exist in the repository, and that these snippets can be efficiently found and exploited~\cite{barr2014fse}. They also reported that, on average, 43\% of changes could be reconstituted from existing code in 15,723 commits from 12 Java projects. In another study, Ray et al.\ considered changes unique if there are no similar or identical lexical and syntactic content, or if they do not undergo the same edit operations, and conducted an empirical study of the uniqueness of changes in the Linux kernel and industrial projects~\cite{ray2015msr}. They further insisted that since there is a considerable number of non-unique changes, developers can be helped in many ways by exploiting those changes. While the above three papers show evidence for repetitive changes, they do not implement tools that can be used to find bugs and fix them. Such empirical studies motivated us to use change patterns to build a tool that could identify buggy code and provide fixes to developers. While each paper comes up with its own way to examine changes, none of them are about a tool~(unlike ours) that can extract changes, abstract them to a pattern, and find buggy code in a given version based on the detected patterns.

\subsection{Change Pattern-Based Approaches}

There are several approaches~(\textsf{FixWizard}~\cite{nguyen2010icse}, \textsf{SBD}~\cite{liang2013esecfse}, \textsf{BugMem}~\cite{kim2006fse}, \textsf{SYDIT} and \textsf{LASE}~\cite{kim2009icse,loh2010icse,meng2011esecfse,meng2013icse}) that can be used to extract patterns from changes or source code snapshots and utilize them to support further changes. These approaches are the ones that are closest to \Tool. However, while they share their motivation with ours, the technical details and the expected outcomes differ from ours. The comparison of \Tool with existing approaches is shown in Table~\ref{tbl:comparison}.

\begin{table}[b]\centering
 \caption{Brief comparison of the bug-fix pattern extraction approaches}\label{tbl:comparison}
 {\begin{tabular}{lllll}\hline
 & \multicolumn{2}{l}{\textit{Cardinality}} & \multicolumn{2}{l}{\textit{Representation}} \\
 Approach & Inputs & Outputs & Bug pattern & Fix pattern \\ \hline
 \textsf{FixWizard} \cite{nguyen2010icse} & One change & One pattern & Program flow graph & Program flow graph \\
 \textsf{SBD} \cite{liang2013esecfse} & One change & One pattern & Graph & Statement Insertion \\
 \textsf{BugMem} \cite{kim2006fse} & One change & One pattern & Token sequence & Token sequence \\
 \textsf{LASE} \cite{meng2013icse} & Changes & One pattern & AST subtree & AST subtree \\
 \Tool & Changes & Patterns & Token sequence & Token sequence \\ \hline
 \end{tabular}}
\end{table}

The most significant difference between the existing approaches and \Tool is that, except for \Tool, all of the other approaches are designed to derive a pattern from changes that have been prepared manually. This means that a developer has to select what changes need to be abstracted to a pattern and then feed them into the approach.  On top of that, approaches like \textsf{FixWizard}, \textsf{SBD}, and \textsf{BugMem} distill only one change instance to a pattern representation that can then be reused. \textsf{LASE}, on the other hand, extracts the commonality of multiple change instances and outputs a change pattern. This means that, to derive a pattern, these approaches require users to manually specify a set of related changes as the source of the derived pattern. In contrast, \Tool extracts change patterns from all the changes and then automatically determines all the PSBPs relevant to a project. This means that the developers do not need to guess which change could potentially be a pattern or code written elsewhere. Since, in all related approaches, the changes had to be curated manually, we are unable to perform meaningful comparisons. In order to prepare all the changes to be fed into the related approaches, it would be necessary to implement another tool. Additionally, even if we were to prepare all the changes manually, we find that, except for \textsf{LASE}, none of the other tools are available. Note that, in the case of \textsf{LASE}, the available version cannot be run with any current version of the Eclipse IDE or Java. Hence, none of the currently available tools can be executed by researchers or developers.

Furthermore, because our approach analyzes all the changes, performance is an important aspect. Although graph-based~(\textsf{FixWizard}~\cite{nguyen2010icse}) or AST-based representations of change patterns are effective when used to precisely express program structures, they require higher computational costs to extract patterns from change instances, which makes them unsuitable when a large number of change instances are used as inputs. Thus, even if we did reimplement all the other tools, they would not scale to repositories with thousands of changes.

\subsection{Other Related Studies}

\textbf{AST differencing.}
AST-based program differencing approaches~\cite{falleri2014ase,fluri2007tse} compare two source code versions, compute tree-edit operations, and then map each tree-edit to atomic AST-level change types. Kim et al.\ proposed an algorithm that identifies entity mapping at the function level across revisions when an entity's name changes~\cite{kim2005wcre}. They also proposed a rule-based program differencing approach that discovers and presents systematic changes as well as high-level software changes~\cite{legoues2015tse}. Although these studies are similar to our approach in that they build tools that distill changes from the repository, they stop at distilling changes and do not conduct evaluations to see if the changes they distilled can be used to fix bugs in any particular version of a project. This is because they do not have a mechanism to match and find latent bugs in a particular version of the project. In contrast, our approach uses the changes and has been used to submit pull requests that have been accepted in real-world projects.

\textbf{Co-change pattern mining.}
\textsf{DynaMine} finds bugfix patterns related to method invocations~\cite{livshits2005esecfse}. 
For example, the tool found that method \textit{writeUnlock} should be invoked after an invocation of method \textit{writeLock} in their experiment.  
If invocations of the methods exist in this order, they are regarded as being used correctly. 
However, if only one of the two methods is invoked, or if the two methods are invoked in the inverse order, such usages are reported as error usage patterns by the tool.
\Tool, on the other hand, does not restrict its analysis to just method invocations and any change can be abstracted to a pattern.

\textbf{Automatic repair.}
Automatic program repair techniques are designed to suggest fixes to developers when a bug is identified~(typically due to a failing test). Typically the fixes are generated through search-based software engineering techniques~\cite{ke2015ase,legoues2012tse}, program synthesis and constraint solving techniques~\cite{long2015esecfse,mechtaev2016icse,nguyen2013icse}, or by manually identifying fix templates in human written fixes. While automated repair focuses on fixing bugs commonly known to humans, our approach will find buggy code automatically, like SA tools, and also suggest possible fixes based on the bug fix history in a project.

\textbf{Pattern mining from source code.}
\textsf{PR-Miner} finds implicit coding rules and detects their violations~\cite{li2005esecfse}. It finds rules with \textit{frequent itemset mining}, which looks for programming elements that frequently occur together in source code. If developers violate rules by failing to include elements that should appear with other elements, \textsf{PR-Miner} can warn them of the problems. Liang et al.\ proposed \textsf{AntMiner}, which improves the precision of mining by removing noise using program slicing~\cite{liang2016icse}. \textsf{MAPO} takes into account the order of program elements by applying \textit{frequent subsequence mining}~\cite{zhong2009ecoop}, which means it can detect order-sensitive problems.

Although code pattern mining techniques can capture coding patterns, they do so in a single snapshot. There is another set of approaches that capture coding patterns in changes. For example, Kagdi et al.\ showed that it was possible to extract the set of files that were changed together from the source code repository and then apply frequent sequence mining to determine which files in that set of files needed to be changed when a particular file undergoes changes~\cite{kagdi2006msr}. 
Zimmermann et al.\ focused on providing much broader granularity for three frequently changing elements: file level, method level, and variable level~\cite{zimmermann2005tse}. 
To accomplish this, they applied association rule mining to guide developers to the elements that need to be changed when a particular element is modified. Hanam et al.\ proposed cross-project bug patterns for JavaScript software~\cite{hanam2016fse} with the goal of discovering the bug patterns that are inherent to JavaScript. However, in contrast to our automated approach, their detection process includes manual work in the component building process. Fluri et al.\ proposed a technique that can be used to find frequent change patterns~\cite{fluri2008ase}, but the technique does not focus on bugfix patterns. 

Code clone detection techniques can also be utilized to find code patterns. For example, Li et al.\ developed a clone detection tool named \textsf{CP-Miner}~\cite{li2006tse} and utilized it to check whether normalized variable names match between clones. If variables in a clone pair are matched partially, it is likely to include a bug that can then be reported to developers. Similarly, Inoue et al.\ applied a code clone detector to two mobile software projects developed in a company and detected 26 latent bugs in the systems~\cite{inoue2012iwsc}. In both studies, inconsistencies, and hence bugs, were identified between clone pairs.

Our approach is fundamentally different from the above approaches in that we mine code changes and not the source code snapshot of a project. Hence, we can see what code is buggy, how to fix that bug, is the change a pattern, are there instances of the buggy code in a given version of the project, and how to potentially fix them.
Although we could have used any of these clone detection techniques in our approach to finding pattern changes, we chose not to because such techniques do not scale well to thousands of changes over thousands of code versions. Our technique, which aims to replace variable names with special tokens and then calculate a hash value for each program statement in order to derive change patterns, is inspired by a few other clone detection techniques~\cite{dang2012acsac,li2006tse,murakami2013icpc,roy2008icpc}.

Overall, we acknowledge that there are clone detection techniques 
and that a variety of SA tools already exist. However, we brought those techniques and tools together, along with change level analysis, to help developers and maintainers find and fix commonly occurring bugs. In the process of doing so, we had to overcome engineering challenges needed to help the tool scale to practical projects and not just toy examples. As an engineering research area, we think that our contributions (bringing previous research ideas together, solving engineering challenges, building a working tool, and conducting a real-world empirical case study with fixed bugs) are highly relevant.

\section{Threats to Validity}
\label{sec:threats}

\subsection{Internal Validity}

Internal validity refers to confounding factors that might affect the causal relations established throughout an experiment~\cite{wohlin2012book}. 
In our experiment, we filtered out the maximum number of false positives possible to ensure the latent bugs identified by our tool would result in a manageable number of pull requests for developers.
Furthermore, while there could also be false positives among these latent bugs, we do not think that this risk is severe because developers can apply the same filtering steps we used in our tool, and thus will not have an excessive number of potential bugs to examine at one time. 
To address any mistakes that could have made in our evaluation or our implementation~(threat to internal validity), we openly provide the source code of our tool, the binary version of our tool, and the raw data collected from applying our tool to the four case study subjects to anyone who would like to examine them.~\footnote{\url{https://doi.org/10.5281/zenodo.3460378}}

There is another risk related to our work. 
Our proposed technique is based on the assumption that the same problematic code will be modified in the same way. 
Thus, if the same problematic code is modified in two or more different ways, our proposed technique cannot detect PSBPs for the problematic code. 
At this moment, however, it is difficult to gauge how often our proposed technique incorrectly filters out PSBPs from change patterns because the number of change patterns is several thousand or more, and it would be unrealistic to manually analyze such a large number of change patterns.
Asking real experts to use the tool is one of our future work.

\subsection{External Validity}

Threats to external validity impact the generalizability of the results obtained in a study~\cite{wohlin2012book}. While we evaluated our tool only on four Java projects that used Git as a version control system, our approach is general enough that it can be applied to any version control system and any programming language. The reason we used the four projects chosen for this study is that they manage issues well with \textsf{Jira}/\textsf{Bugzilla}, which meant we were able to easily obtain the IDs of the resolved and closed bug-related issues. Our approach utilizes bug-related issue IDs to determine whether or not a given commit is a bug fix. More specifically, if a log message of a given commit includes any of the bug-related issue IDs, it is regarded as a bug fix commit. For example, in the case of \textsf{Camel}, the bug-related issue IDs are "CAMEL-72" or "CAMEL-80". We believe that our method of using bug-related issue IDs is equal to or better than methods that use keywords such as "bug" or "fix" to identify bug fix commits. We also manually confirmed that the 19 single-match PSBPs consists of at least a bug-fix commit.
Note that the PSBP extraction approach still works if clean bug-fix data does not exist. However, we think that there would be more false positives as non-bug-fix commits might be included in the analysis.

\subsection{Construct Validity}

Construct validity refers to the degree to which the various performance measures accurately capture the concepts they intend to measure~\cite{wohlin2012book}. In our experiment, there were minimal threats to construct validity since we evaluated the proposed technique by using the number of pull requests that were accepted by the developers of the target projects.

\section{Conclusions}
\label{sec:conclusion}

In this paper, we proposed a new technique named \Tool\ to identify project-specific bug patterns (PSBPs).  We derive those PSBPs from the past development history of a given software project and use them to find latent buggy code. Our proposed approach not only finds buggy code in a given revision of a software project, it also suggests a solution for each buggy code that is identified. We also implemented a software tool based on our proposed approach and applied it to four open source software projects.  In doing so, we brought together previous research ideas and overcame engineering challenges that helped the tool scale up to practical projects and not just toy examples.  Our evaluation indicates that our tool was useful for identifying latent buggy code in a given revision of a software project. Indeed, five out of the six pull requests that we made based on our tool's findings were merged by the developers of their related software projects. Furthermore, our analysis of the false positives identified in this study can be expected to provide us with guidance on how we can improve our approach and tools in the future.

\end{document}